%Paper: chao-dyn/9309002
%From: njb@phys.columbia.edu (Neil J.Balmforth)
%Date: Wed, 8 Sep 93 11:06:43 EDT

% New York, August 31, 1993
%\magnification=1100
%\nopagenumbers
 \hfuzz=6pt
%\raggedbottom
%\headline={\ifnum\pageno>1\hss -- \tenrm\folio -- \hss
%\else\hfil\fi}

\def\mb#1{\setbox0=\hbox{#1}\kern-.025em\copy0\kern-\wd0
\kern-0.05em\copy0\kern-\wd0\kern-.025em\raise.0233em\box0}

\font\ninerm=amr9
\font\bfs=ambx9
\font\its=amsl9

%\font\title=mvb10 scaled \magstep2
 \font\title=ambx8 scaled \magstep2
%\font\title=cmbx10 scaled \magstep2
%\baselineskip 24pt

%\baselineskip 17.5pt

\def\bfxi{{\mb{$\xi$}}}
\def\ltwid{\mathrel{\raise.3ex\hbox{$<$\kern-.75em\lower1ex\hbox{$\sim$}}}}
\def\gtwid{\mathrel{\raise.3ex\hbox{$>$\kern-.75em\lower1ex\hbox{$\sim$}}}}
\def\c{\centerline}
\def\bfx{{\bf x}}
\def\bfz{{\bf 0}}

\def\d{\dot}
\def\dd{\ddot}
\def\e{\eqno(}

\def\ddd{\raise 6pt\hbox{\hskip .3pt {.}\kern -.8pt{.}\kern -.8pt{.}}}
\def\r{\rlap}
\def\up{{\uparrow}}
\def\do{{\downarrow}}

%e.g. $\rlap {\ddd} x $

\def\cn{{f}}
\def\clr{{\cal R}}
\def\cl{{\cal L}}
\def\ve{{\varepsilon}}
\def\b{\bigskip}
\def\cls{{\cal S}}
\def\cln{{\cal I}}

\c {{\title CHAOTIC PULSE TRAINS}\footnote{$^*$}{ %
This work was supported by the U.S. Air Force under %
contract number F49620-92-J-0061, and by a post-doctoral fellowship %
to NJB from the Science and Engineering Research Council of England and Wales}}

\bigskip

\c {N. J. BALMFORTH\footnote{$^\dagger$}{ %
Astronomy Department, Columbia University, New York, NY 10027},
G. R. IERLEY\footnote{$^\ddagger$}{ %
Scripps Institute of Oceanography, University of California, San Diego, %
CA 92093-0225} {\ninerm AND} E. A. SPIEGEL$^\dagger$}

\bigskip

{\bfs Abstract.}
{\ninerm We study a third-order nonlinear ordinary differential equation whose
solutions, under certain specific conditions, are individual pulses.
These correspond to
homoclinic orbits in the phase space of the equation and we
study the possible pulse types in some detail.  Sufficiently close to
the conditions under which a homoclinic orbit exists, the solutions
take the form of trains of well-separated pulses. A measure of
closeness to homoclinic conditions provides a small parameter for the
development of an asymptotic solution consisting of superposed,
isolated pulses.  The solvability condition in the resulting singular
perturbation theory is a {\its timing map} relating successive pulse
spacings.  This map of the real line onto itself, together with the
known form of the homoclinic orbit, provides a concise and accurate
solution of the equation.

\medskip

{\bfs Key words.}
Ordinary differential equations, maps,
homoclinic orbits, chaos, nonlinear
dynamics, singular perturbation theory.

\medskip

{\bfs AMS(MOS) subject classifications.} 34A45, 34C35, 34C37, 34E05,
34E15, 58F31.}

\bigskip

{\bf 1. Introduction.}
Temporal sequences of pulses are a common feature of nonlinear
oscillators.  In this paper we consider one such oscillator whose
amplitude, $x(t)$, is described by the ordinary differential equation (ODE),
$$
\r {\ddd} x +  \mu \dd x + \d x - c x + \cn (x) = 0,
\e1.1)
$$
where $f = x^n$, with $n=2$ or 3, the dot implies differentiation with
respect to the argument, and $\mu$ and $c$ are parameters.  Such
equations arise in problems with competing instabilities [2],
[3], [26]. For certain parameter values and with appropriate initial
conditions, the solutions take the form of irregularly spaced pulse
trains.

Trains of propagating pulses, or coherent structures, are also
encountered as solutions to partial differential equations [6], [12].
For integrable systems, these are composed of solitary waves and can
be studied exactly, if somewhat convolutedly, with inverse scattering
transforms [1]. Here we are interested in dissipative systems for
which no such exact, general theory exists.  An example is Benney's
equation [5],
$$
\partial_T u + \partial_X u^n =
- \partial_X^2    u
- \mu\partial_X^3 u
- \partial_X^4    u
\quad, \eqno(1.2)
$$
where $T$ and $X$ are the time and space variables.
This equation (for $n=2$) was found in the study of instability in thin liquid
films. More importantly, it describes phase evolution for the
complex Ginzburg-Landau equation in a general reference frame,
and so our considerations apply to many spatially extended systems
subject to Hopf bifurcation.
Numerically, Benney's equation has been observed to produce pulses over
wide ranges in the dispersion parameter, $\mu$ [11], [21].  It
integrates to an ODE on introduction of a traveling-wave
solution, $x(t)=u(X-cT)$, where $c$ is the wave speed. Here we consider
a special case of that ODE, namely
equation (1.1), when the constant of integration vanishes.

Solitary waves, or propagating pulses, in the dissipative system (1.2)
correspond to homoclinic orbits in the phase space, $\bfx=(x,\dot
x,\ddot x)$, of equation (1.1).  This association indicates that
methods used to construct solutions of partial
differential equations consisting of several solitary waves may also
be used to build pulse trains [8]. Those methods were conceived in
nonlinear field theory to describe the equations of motion of
particles [9], and have been extended to nonlinear dissipative systems
like (1.2) when there exist solitary waves with a modicum of stability
[18], [22], [29].  If the pulses of a train are well separated, the
solution consists approximately of a superposition of homoclinic
orbits [17]. Guided by the solitary-wave techniques, we use this
approximate solution as a basis about which we develop a singular
perturbation expansion for (1.1). This asymptotic analysis then
provides an analytical solution for pulse trains once the pulse form is
determined.

We organize the paper as follows. In sections 2-4, we describe the
homoclinic orbits and pulse-train solutions of equation (1.1) in some
detail.  In section 5, we develop the asymptotic method to first
order, and derive the timing map for the pulse separations.  In
section 6, we generalize the asymptotic theory slightly to accommodate
a symmetry of the equation with cubic nonlinearity (the case $n=3$),
and in section 7 we discuss higher-order asymptotic corrections for
this case. In section 8 we deal with some numerical tests of the
asymptotics, and consider some special pulse-train
solutions in section 9.
In section 10, we explain our failure to deal adequately
with the incompressible limit of equation (1.1), and in section 11 we
offer some final remarks. In Appendix A we give further computational
details and numerical accuracy checks.

\goodbreak
\b
{\bf 2. Orbits and pulses.}

{\bf 2.1 Homoclinic orbits.}
When the system is in the neighborhood of the origin of
its phase space, ${\bf x} = (x,\dot x, \ddot x)={\bf 0}$,
equation (1.1) is well approximated by its local linearization:
$$
{d {\bf x}\over dt}  = \
\pmatrix{0 & 1 & 0\cr 0 & 0 & 1 \cr c & -1 & -\mu \cr} \bfx
\equiv A \bfx
\quad.
\eqno(2.1)
$$
This equation has the solution,
$$
\bfx = \bfxi_1 e^{s_1t} + \bfxi_3 e^{s_3t} + \bfxi_3 e^{s_3t}
\quad,
\eqno(2.2)
$$
where the constant vectors, $\bfxi_i,$ are the eigenvectors of $A$,
and its eigenvalues, $s_i$, satisfy the cubic equation,
$$
s^3 + \mu s^2 + s - c = 0
\quad.
\eqno(2.3)
$$
In this region of phase space, the behavior of solutions is dictated by
the eigenvalues, $s_i$, and the eigenvectors, $\bfxi_i$.  The latter
span the stable and unstable manifolds which intersect at the origin
and whose geometry critically controls the local dynamics.

For the flows we discuss here, the parameters $\mu$ and $c$
are chosen such that among the three eigenvalues there is one that is
real and positive, $s_1=\gamma$, while the other two form a complex
conjugate pair with negative real part, $s_{2,3}=-\sigma\pm i\omega$
(so $\gamma$, $\sigma$ and $\omega$ are all real and positive).

In addition to the origin, we also have {\it secondary fixed points} that are
located on the $x-$axis at
$$
x_0 =
\cn(x_0)/c, \qquad x_0\ne 0
\quad.
 \eqno(2.4)
$$
In the cases we consider here, with $f=x^n$, we have $x_0 = c$ for
$n=2$, and $x_0 = \pm \sqrt{c}$ for $n=3$.

For such flows, a trajectory starting from the neighborhood of the
origin on the unstable manifold, departs exponentially quickly along
the direction of the unstable eigenvector. For a range values of $\mu$
and $c$, this trajectory then loops around one of the secondary fixed
points and spirals back in towards $\bfx = {\bf 0}$.  In fact, if
these parameters are suitably tuned, the trajectory enters the stable
manifold spanned by the other two eigenvectors and ultimately winds
into the origin.  In that case, the system has a {\it homoclinic
orbit} which is biasymptotic to $\bfx = {\bf 0}$.  (For some dynamical
systems, the existence of the homoclinic connection of the unstable
and stable manifolds has been proved. Here, though, we speak
qualitatively and, in saying that a certain solution exists, we
typically mean that we have found it by numerical methods.)

\goodbreak

\medskip
{\bf 2.2 Pulse trains.}
Homoclinic orbits exist only at certain points $(\mu_0,c_0)$ on
special {\it loci} in the parameter space $(\mu,c)$.  When the system
is on such a locus, we say that it is in a homoclinic condition.
Typically, such a condition is not robust; a small generic change in
parameters will destroy the homoclinic condition. Instead, in these
nearly homoclinic conditions, one frequently finds a multitude of
orbits of finite period that to some extent approximate the homoclinic
original.  The union of these orbits comprises a complicated set, but
if it attracts trajectories for extensive periods of time, then the
system executes a series of pulses and can create irregularly spaced
trains.

In order for the set to temporarily capture trajectories, the periodic
orbits must have some degree of attractiveness. In terms of Floquet
exponents, all of the orbits must have one exponent that is large and
negative, signifying that the flow contracts regions of phase space
towards the set.  Of the remaining two exponents, one is always zero
since it corresponds to motion along an orbit. The third describes the
stability of each orbit within the strange set.  In fact, when the
Shil'nikov parameter,
$$
\delta = \sigma / \gamma
\quad,
\eqno(2.5)
$$
is less than unity, most orbits become unstable within the set [27],
[28], [29], and this situation is commonly used to predict the
existence of chaotic dynamics.

When the system is in a nearly homoclinic condition, we set
$$
\mu=\mu_0+\ve \mu_1
\qquad {\rm and} \qquad
c=c_0+\ve c_1
\quad,
\eqno(2.6)
$$
where $\ve$ is a small parameter to be defined later and $\mu_1$ and
$c_1$ are $O(1)$.  For these values of $\mu$ and $c$, the trajectory
follows a path that shadows the original homoclinic orbit. However, it
typically winds near the stable manifold without intersecting it and
cannot therefore spiral directly into the origin on returning to its
vicinity. Instead it successively misses $\bfx = {\bf 0}$ by
distances of the order of $\ve$, and executes multiple loops around the
other fixed point. The outcome is a temporal sequence of pulses.

Pulse trains of this kind are shown in Fig. 1.  They were generated
numerically from equation (1.1), beginning from an initial condition
just off the origin and on the unstable manifold.  Panel (a) shows a
pulse train computed for $n=2$ with $\mu=1/\sqrt 2$ and $c=1.92847$.
Panel (c) displays a pulse train for the cubic equation with $\mu =
1/\sqrt 3$ and $c=1.04430$.  These values ensure conditions in which the
leading pulse is followed by a second, similar one; thereafter, the
timings of the pulses becomes more complicated.  In the quadratic
case, after a few, nearly homoclinic pulses, the trajectory eventually
finds its way around the stable manifold intersecting the origin and
diverges to $-\infty$.  The cubic case has a richer behavior because
this equation has a fixed point on the negative axis. On escaping
around the stable manifold, the trajectory loops this negative fixed
point and returns to the neighborhood of the origin, so creating an
``antipulse.'' Pulse trains in the cubic case consequently contain a
mixture of pulses and antipulses, with reversals in ``polarity''
occurring where the solutions for the quadratic case would have
diverged.

\b
\goodbreak
{\bf 3. Orbital classification.}

{\bf 3.1 Order.}
The pulse trains shown in Fig. 1 are related to one particular
variety of the homoclinic orbits admitted by equation (1.1).  We call
such an orbit a {\it principal homoclinic orbit} because, on leaving
the origin, it loops only once around another fixed point before
completing its circuit by spiralling back into $\bfx={\bf 0}$. In
later sections we consider pulse trains made up of solely principal
homoclinic orbits. But equation (1.1) admits many other kinds of
homoclinic orbits from which trains could similarly be composed, and
we provide a brief description of these various pulse types.

Homoclinic connections can also be formed by trajectories that
encircle the other fixed point more than once before spiraling back
into the origin. Homoclinic orbits with two loops have been found
previously for (1.1) [2] and have been called
{\sl secondary} orbits [15]. More generally, such homoclinic orbits can be
roughly classified according to the number, $N$, of times the
trajectory loops around the secondary focus before it is reinjected
back into the neighborhood of the origin.

This notation is meant simply to provide an informal way of referring
to solutions. Moreover it only covers a small, fundamental subset of
all homoclinic orbits. The elements of this subset we
refer to as {\sl basic orbits}; they pass through the neighborhood of
the origin only at the beginning and the end of the trajectory.  A
multitude of other homoclinic connections are achieved when the
trajectory is permitted to pass through this neighborhood an arbitrary
number of times. Because every pass by the origin is comparatively
slow and amplitudes remain small there, these connections may be
thought of as pairings, triplings and so on of the single
pulses.  These {\sl composite orbits} are numerous and densely
populate the vicinity of the homoclinic loci of the basic orbits.  In
effect, they are homoclinic pulse trains.

In the asymptotic analysis of section 5, pulse trains are constructed
as superpositions of basic orbits. Consequently the composite
homoclinic orbits can be entirely described within the framework of
the asymptotic theory and we do not discuss them further.

\goodbreak
\medskip
{\bf 3.2 Polarity.}
The foregoing discussion of the order $N$ of a homoclinic connection
ignores the possibility that there may be more than one fixed point
around which trajectories can loop before spiraling into the origin.
It is therefore restricted to the quadratic case. In the cubic case,
there are two fixed points away from the origin and so there are two
families of basic orbits mirroring each other.  With a member of each
family, one can again associate composite homoclinic orbits, but there
is a greater variety of such compositions because connections can
contain both pulses and antipulses.

A more unusual kind of orbit generated by the cubic equation is a
multiply peaked pulse with polarity reversals within the pulse
itself.  This is a new kind of basic orbit which can be placed into a
modification of the classification scheme if we record the relative
polarity of the peaks in the pulse, in addition to the total number of
turns, $N$, around both secondary foci.  Thus, for example, there are
two types of $N=2$, doubly peaked pulses; a $(\up \up)$ pulse and a
$(\up \do)$ pulse (plus their reflections),
of which only the former has a counterpart in the
quadratic case.

\goodbreak
\b
{\bf 4. Homoclinic loci.}

{\bf 4.1 Quadratic case.}
Fig. 2 displays the locus for the basic homoclinic orbits in the
$\mu-c$ plane for the quadratic case.  The branch of principal orbits
descends from $(\mu,c) = (\infty,\infty)$, and crosses the axis
$\mu=0$, which corresponds to the dispersionless special case of the
Benney equation, and is known as the
LaQuey-Mihajan equation [25] or the Kuramoto-Sivashinsky equation
[24].  Shortly beyond this point, the locus sharply bends around and
turns into a secondary homoclinic branch.  This locus winds first to
positive, then to negative $\mu$ where it bends sharply around and
turns into the tertiary branch.  There follows a sequence of such
winding branches, of which only the first three are shown in the
figure.  However, the pattern continues through many successive
winding branches until the locus ends at the origin.  Throughout this
cascade, the pulses acquire more and more, tighter and tighter loops
around the secondary focus, until at $\mu=c=0$ the two fixed points
merge. A more complete study of the homoclinic orbits for the
quadratic case (and also various periodic solutions)
is offered by Chang, Demekhin and Kopelevich [7].

In order for the homoclinic branches to transform into one another,
there must be distinct topological changes in the orbit at certain
bifurcation points on the loci. We have not isolated these points nor
followed the development of these bifurcations. Instead, we have
generated the loci portrayed in the figure (and also their cubic
counterparts shown in Fig. 3) as sequences of solutions of a
boundary-value problem.  For convenience, these were periodic on very
large domains (see Appendix A). Other methods gave identical solutions
at various points on the loci, but we cannot exclude the remote
possibility that the loci are more complicated near the bifurcation
points.

\medskip
\goodbreak
{\bf 4.2 Cubic case.}
The loci of some of the lowest-order homoclinic orbits are shown in
Fig. 3.  The behavior of the principal branch is, at least
initially, similar to the quadratic case. At its terminus, however,
the locus bends upwards into the second-order homoclinic branch with
polarity $(\up \do)$.  This secondary branch then spirals in towards
the point $(\mu,c) \simeq (0.67,-1.28)$, achieving higher order on
each left-hand turn of the spiral.  That spiral point corresponds to a
heteroclinic connection of the origin to another fixed point, with a
single loop around the other focus (see the final phase portrait of
Fig. 5).  Heteroclinic connections of this kind are found at
distinct points in parameter space, rather than loci, because these
orbits necessarily join the one-dimensional invariant manifolds of two
fixed points. This can only be achieved at unique conjunctions,
or co-dimension two bifurcations. A similar picture emerges for the
Lorenz system [16].
(There are various other heteroclinic connections admitted in the
phase space that we have not considered at all. These connect to the
two-dimensional invariant manifolds of one of the
secondary fixed points and
follow {\sl heteroclinic loci}.)

The topologies of the loci of the $N=2 \; (\up \up)$ and the $N=3 \;
(\up \up \up)$ orbits are also different in the cubic case.  These no
longer join continuously to one another and to the principal branch.
Instead, the winding locus is broken apart and the individual pieces
self-connected by curves corresponding to one of the mixed-polarity
basic orbits. These orbits are like the original pulses, but with an
added polarity-reversed peak.  Thus, for example, the $N=2 \; (\up
\up)$ branch becomes self-connected to the left by the locus of the
$N=3 \; (\up \up \do)$ orbit, creating a closed contour in the $c-\mu$
plane (see Fig. 3).  Inside the closed contours, the loci of orbits
with even more reversed peaks are nested.
%and each pattern surrounds a heteroclinic point.
These nests of contours trail off towards the
origin, where, as in the quadratic case, all the fixed points merge.

\medskip
\goodbreak
{\bf 4.3 Orbital morphology and pulse train phenomenology.}
The shapes of nearly homoclinic trajectories for a sampling of
parameter values near to the winding homoclinic loci of the quadratic
case are shown in Figs. 4 and 5. On the loci at larger values of $\mu$,
the spiraling of the trajectory into ${\bf x}=0$ is
rapid, whereas at smaller $\mu$ it is slow. When one homoclinic branch
transforms into one of higher order, the orbit apparently captures a
loop of the decaying spiral into the origin.

The shapes of the orbits are connected to the dependence of the linear
eigenvalues on $\mu$ and $c$.  This parametric dependence also has
repercussions on the form of the pulse trains. Where the orbit spirals
back into the origin relatively rapidly, indefinitely long pulse
trains with constant pulse spacings tend to form. On the left of the
$\mu-c$ plane, where the decay is slow, the trajectory rapidly departs
from the homoclinic orbit, and trains of more than one pulse are very
difficult to find.  In between, the pulse spacings vary in a complex
fashion, sometimes chaotically, and trains invariably involve a finite
number of pulses.

Surprisingly, this is true even for the cubic case, which one might
have thought always generated indefinitely long trains, since the
solution avoids diverging beyond the stable manifold by executing an
antipulse.  In fact, on executing a pulse, the trajectory typically
migrates further and further from the origin. Consequently, spacings
become smaller and smaller, until eventually they are too short
meaningfully to isolate individual pulses.
Once it has escaped from the vicinity of the origin entirely,
the trajectory spirals ever farther outwards on a diverging
path. Thus, in spite of the generally stabilizing effect of cubic
nonlinearity, trains do not continue indefinitely but terminate as
pulses lose distinguishability through strong overlap and a ``spacing
catastrophe'' occurs. We could avoid this behavior by including
further nonlinear terms in the equation, such as $x^2\dd x$, but the
topology of loci in the $(c-\mu)$ plane would then be very different.

Pulse train solutions exist only very near to the homoclinic loci, and
only in very narrow bands whose width depends on the location on the
locus.  From Fig. 2 we can anticipate where the bands are; their
thicknesses must be judged by other means. The pulses within the train
can be unambiguously resolved only when their separations are larger
than the intrinsic pulse width. The separations themselves shorten as
one tracks away from homoclinicity, and so we can gauge how far one can move
in $\mu$ and $c$ from the homoclinic values $(\mu_0,c_0)$ before
pulses become obscured. This sets the thicknesses of the bands
surrounding the homoclinic loci in which we expect to find pulsatile
solutions that can be analyzed by the asymptotic theory.  For the two
pulse trains shown in Fig. 1, the approximate
range in $c$ at fixed $\mu$ is one percent,
which bounds $|\ve c_1|$ from above. For larger $\mu$, the
bands are wider, for smaller $\mu$ the bands become very thin.

\goodbreak
\b
{\bf 5. Pulse dynamics.}

{\bf 5.1 The Ansatz.}
In the limit that the pulses of a train are well resolved and widely
separated, a solution can be extracted from the ODE by asymptotic
means.  For sufficiently large separations, the pulses interact weakly
and only mildly distort each other, and so the solution is
approximately a linear superposition of single-pulse solutions, or
homoclinic orbits.  At infinite separation, such a linear
superposition reduces to an isolated homoclinic orbit and is therefore
exact. This is the basic solution with which we open a singular
perturbation expansion, but first we need a suitable expansion
parameter.

In section 2, we introduced $\ve c_1 = c-c_0$ as a measure of
parametric proximity to the homoclinic locus. When thinking
of pulse interactions, another convenient small parameter is
one that measures the amplitude of one pulse at the position
of its nearest neighbors.  However, this is not an imposed
parameter, but is implicitly fixed by our choice of $\mu$
and $c$. By identifying it with $\ve$, we obtain a useful and
consistent asymptotic expansion.

The amplitude of a pulse outside of its core is set by the exponential
decay rates fore and aft, or equivalently, by the eigenvalues,
$\sigma$ and $\gamma$, of linear theory.  If the characteristic pulse
separation is $\bar{\ \Delta \ }$, then we set
$$
\ve = \exp(-\gamma \bar{\ \Delta \ })
\quad,
\eqno (5.1)
$$
and so $\ve$ measures the amplitude of a pulse in the vicinity of its
predecessor.  Near the position of its successor, the amplitude is
$\ve^\delta$, where $\delta=\sigma/\gamma$ is the Shil'nikov parameter
of \S 2.2.  Here we concentrate chiefly on situations with
$\delta\simeq1$, for which pulses decay at similar rates fore and aft.
This places us in the vicinity of the curve $c=2\mu^3+\mu$
in parameter space, which is drawn on Figs. 2 and 3. In a later section,
we briefly consider trains of asymmetric pulses for
which $\delta$ takes values as small as 1/2.

For the homoclinic orbit we write $x(t) = H(t)$ and choose the phase
of $H$ so that its principal maximum occurs at $t=0$. The weak
interaction approximation is expressed in the ansatz,
$$
x(t) = \sum_{k=1}^K H(t-\tau_k) +\ve \clr(t, \{\tau_m\},\ve)
\quad.
\eqno(5.2)
$$
The main peaks in this pulse train are at $t=\tau_k, \ k=1,\dots,K$,
where $K$ may be infinite.

To compensate for the error in the assumption of linear superposition,
we have included the correction, $\ve \clr(t, \{\tau_m\},\ve)$.  This
error term must vanish as the pulse separation becomes infinite and
$\ve \rightarrow 0$ so, by insisting that $\clr(t, \{\tau_m\},\ve) $
remain finite, we are led to a solvability condition in the manner of
singular perturbation theory.  That condition constrains our choice of
the set, $\{\tau_m\}$, and this, with equation (5.2), provides an
explicit solution up to order $\ve$.  We further reduce the error by
solving the equation determining $\clr(t, \{\tau_m\})$ in section 7,
but even without that improvement, the leading-order approximation is
adequate for many purposes at small $\ve$.

\bigskip
\goodbreak
{\bf 5.2 First-order perturbation theory.}
The homoclinic solutions, $H(t-\tau_k)\equiv H_k$, satisfy the equation,
$$
{d^3\ \over dt^3} H_k + \mu_0 {d^2\ \over dt^2} H_k
+ {d\ \over dt} H_k  - c_0 H_k + (H_k)^n =0
\quad.
\eqno(5.3)
$$
This is the leading-order equation which one obtains on substituting
the ansatz (5.2) into the ODE (1.1) and gathering together terms of
order unity at each $k$.
The corrections that appear at first-order in $\ve$
satisfy the equation,
$$ \ve \cl \clr = \ve \sum_{k=1}^K \left(
 c_1 H_k - \mu_1 \dd H_k  \right)
 - \left(\sum_{k=1}^K H_k\right)^{n}
 + \sum_{k=1}^K (H_k)^n + O(\ve^2)
\quad ,
\eqno(5.4)
$$
where
$$
\cl  = {d^3\ \over dt^3} + \mu_0 {d^2\ \over dt^2} + {d\ \over dt}
- c_0 + n \left(\sum_{k=1}^K H_k \right)^{n-1}
\quad.
\eqno(5.5)
$$

To reduce equation (5.5) still further, we make the following
observation. The operator $\cl$ is like the Schr\"odinger operator for
an electron in a crystal lattice in that it has a potential which is
sharply peaked in certain discrete locations. In the vicinity of these
lattice sites, $\cl$ is approximately given by the local operator,
$$
 \cl_k  = {d^3\ \over dt^3} + \mu_0 {d^2\ \over dt^2} +
{d\ \over dt}  - c_0 + n H_k^{n-1}
 \quad,
 \eqno(5.6)
$$
which has a set of eigenfunctions that are peaked at $t=\tau_k$. In
particular, it has the null vector $\dot H_k$, as one can see by
differentiating equation (5.3).

We define the operator adjoint to $\cl_k$ according to the relation,
$$
\int_{-\infty}^{+\infty}X(t)\cl_k^{\dagger}Y(t) dt =
\int_{-\infty}^{+\infty} Y(t) \cl_k  X(t) dt
\quad,
\eqno(5.7)
$$
for any differentiable
functions, $X(t)$ and $Y(t)$, that vanish at $t=\pm \infty$.
Although we cannot prove that the adjoint operator possesses a
null vector, $N_k(t)=N(t-\tau_k)$, satisfying,
$$
\cl_k^{\dagger} N_k(t) = 0
\quad,
\eqno(5.8)
$$
we have numerically found one to exist.
Far from $t=\tau_k$, this null vector
behaves like $H(\tau_k-t)$, which decays exponentially and
reveals $N_k$ to be localized to the vicinity of
the $k^{th}$ peak. Therefore, on multiplying the right-hand side of
equation (5.4) by this vector and integrating, we see that, in view of
(5.8),
$$
\int_{-\infty}^\infty N_k \cl \clr dt = \int_{-\infty}^\infty N_k \cl_k
\clr dt + O(\ve) = O(\ve)
\quad.
\eqno(5.9)
$$
When applied to both sides of (5.4), this gives
$$
c_1 \int_{-\infty}^\infty H_k N_k dt  - \mu_1
 \int_{-\infty}^\infty  \dd H_k N_k dt
-
{1\over \ve} \int_{-\infty}^\infty n (H_k)^{n-1}
 [H_{k+1} + H_{k-1}]
 N_k dt = O(\ve)
 \quad.
 \eqno(5.10)
$$
This solvability condition on the leading-order solution
determines the peak locations, $\tau_k$, in the form of a {\sl timing map}.

\medskip
\goodbreak
{\bf 5.3 The timing map.}
We denote the interval between successive pulses by,
$\Delta_k=\tau_k-\tau_{k-1}$, and (as in
[11]) define the function,
$$
F(\Delta) = {n\over\ve I_0}
\int_{-\infty}^\infty N(t) [H(t)]^{n-1} H(t+\Delta) dt
\quad,
\eqno(5.11)
$$
where
$$
I_m = \int_{-\infty}^\infty N(t) {d^m\over dt^m} H(t) dt
\quad.
\eqno(5.12)
$$
Equation (5.10) can then be written (without the order $\ve$ terms)
in the form,
$$
C_1 = F(-\Delta_{k+1}) + F(\Delta_k)
\quad,
\eqno(5.13)
$$
where
$$
C_1 = c_1 - \mu_1{I_2\over I_0}
\quad.
\e5.14)
$$
This is a relation between adjacent intervals.

For the homoclinic orbits discussed in Section 3, $F(-\Delta)$ decays
monotonically with $\Delta$ as long as $\Delta$ is larger than about $2/10$
of the width of a pulse. Over the range of spacings of interest,
$F(-\Delta)$ can therefore be inverted. The map can then be written in a
more convenient form in terms of the function $Z_k=-F(-\Delta_k)$:
$$
Z_{k+1} = f(Z_k) - C_1
\quad,
\eqno(5.15)
$$
where $f(Z)=F[-F^{-1}(-Z)]$.  Equation (5.13) defines a map of the real
line onto itself, and we use $Z_k$ to portray many of our results.
However, rather than use $Z_k$, we do not scale out the
divisor $\ve$ of the integral in $F(\Delta)$ (which would involve
specifying explicitly one of $\ve$ and $C_1$), and
instead we plot $\ve Z_k$,
which is a more natural unit.

The timing map (5.13) is the basic result of the asymptotic theory.
Its main ingredient is the function $F(\Delta)$, an example of which
is reproduced in Fig. 6. The map generated by this function, shown
in Fig. 7 in both spacing and $Z_k$, describes pulse spacings for
the quadratically nonlinear equation. When defined in terms of
spacings, the map incompletely covers the interval since the relation
(5.13) cannot always be solved for $\Delta_{k+1}$
(specifically when $C_1-F(\Delta_k)>0$, since according to Fig. 6,
$F(-\Delta_{k+1})$ cannot be positive). The spacings for
which no solution exists correspond to the negative branches of the
$Z_k$-map, and they are bracketed by divergences in the function
$\Delta_{k+1}(\Delta_k)$.

To construct $F(\Delta)$ we use the fact that the homoclinic solution,
$H(t)$, is well approximated by a periodic orbit computed on a very
large domain with Chebyshev expansions. Accuracy can be augmented
by using asymptotic tails for $t\rightarrow\pm\infty$ (further
details are given in the Appendix A).
The calculation also provides the
length of the periodic orbits as a function of the distance to
homoclinicity, $c-c_0$.  A sample of the periods are shown in
Fig. 8.  These follow a path equivalent to the bifurcation diagram
shown in Glendinning and Sparrow [15], and point to the infinite number of
periodic orbits shed by the principal homoclinic orbit predicted by
Shil'nikov's theory. From Fig. 8 one can read off the value of
$c-c_0$ that generates orbits with period comparable to the intrinsic
pulse width (about 10 in this case).  That value is about 0.02
(or one percent)
and was quoted earlier as the distance one can move in $c$, at fixed
$\mu$, from $c_0$ before the solution no longer resembles a sequence of
pulses.

\medskip
\goodbreak
{\bf 5.4 The limit of large separation.}
Over a large range in $\Delta$, $F$ is mainly determined by the
exponential, asymptotic behavior of $H(\Delta)$.  For the homoclinic
orbits we are interested in, that behavior is
$$
H(\Delta) \sim e^{-\sigma \Delta} \cos (\omega \Delta + \phi)
\quad,
\eqno(5.16)
$$
and
$$
H(-\Delta) \sim e^{-\gamma \Delta}
\quad,
\eqno(5.17)
$$
where $\Delta$ is assumed positive and $\phi$ is a constant phase.
With these forms for the homoclinic solution, the dependence of the
function $F$ on $\Delta$ factors out of the integral (5.11), and the
map reduces to the Shil'nikov map:
$$
Z_{k+1} \sim A Z_k^\delta \cos[\omega \log Z_k/\gamma - \psi] - B
\eqno(5.18)
$$
[8], for some constants $A$, $B$ and $\psi$.  An important parameter
of this relation is the Shil'nikov parameter $\delta$. This parameter
controls the mean curvature of the map, $Z_{n+1}=f(Z_n)-C_1$.  When
$\delta<1$, the map has a concave envelope, and iterations tend to
diverge quickly from initial locations towards larger values of $Z$,
creating a spacing catastrophe as mentioned above.  In the other
limit, the envelope is convex and iterations typically converge to fixed
points and limit cycles of the map, which correspond to pulse trains
with constant or periodically varying spacings.  This qualitatively
explains the parametric behavior of pulse trains described in
section 4.3.

\bigskip
\goodbreak
{\bf 6. Polarity reversal.}
For simplicity, we have so far only considered the theory for trains
of identical pulses, which is generic for the quadratically nonlinear
ODE. The cubic equation, on the other hand, also admits pulse-train
solutions that are comprised of both pulses and antipulses. We can
incorporate this richer behavior into the analysis by adding a
parameter, $\theta_k$, specifying the polarity, $\pm 1$, of the
$k^{th}$ pulse.  The ansatz is then modified to:
$$
x(t) = \sum_{k=1}^K \theta_k H(t-\tau_k) +\ve \clr
\quad.
\eqno(6.1)
$$
One can once more perform the asymptotic expansion of the solution to
derive the solvability condition on the remainder term $\clr$. The
timing map for this case is
$$
C_1 = \Theta_{k+1}  F(-\Delta_{k+1}) +
      \Theta_k  F(\Delta_k)  \ .
\eqno(6.2)
$$
where $\Theta_k = \theta_k\theta_{k-1}$. Again, we introduce the
iterative variable $Z_k$ and the map becomes
$$
\Theta_{k+1} Z_{k+1} =  \Theta_k f(Z_k) - C_1
\eqno(6.3)
$$
({\it cf.} [14]).  Once the iteration has proceeded from the initial
pair of pulses, the right-hand side of this expression is known at any
stage.  The current spacing, and the relative polarity of the two
pulses separated by this amount, are both given. To iterate the map,
one simply computes the right-hand side; the magnitude determines the
subsequent spacing, and the sign determines whether the next pulse
flips polarity.

To view the process as a simple one-dimensional map, we take as
coordinates $(\Theta_{k} Z_{k},\Theta_{k+1} Z_{k+1})$.  The map then
corresponds to two curves defined by
$$
\Theta_{k+1} Z_{k+1} = f(Z_k) - C_1  \qquad {\rm in} \qquad
\Theta_k Z_k > 0,
\quad,
\eqno(6.4)
$$
which yields a duplication map, and
$$
\Theta_{k+1} Z_{k+1} = - f(Z_k) - C_1 \qquad {\rm in} \qquad
\Theta_k Z_k < 0
\quad,
\eqno(6.5)
$$
providing a reversal map.  An example of both maps is shown in Fig.
9.  This figure has two panels and shows the map in
spacing and $\Theta_kZ_k$.
The spacing map formally diverges whenever $Z_{k+1}$
passes through zero, and is double valued (this nonuniqueness
could be avoided by
dividing the map into two pieces and placing them in different parts
of the $\Delta_k-\Delta_{k+1}$ plane, which is equivalent to using a
variable like $\Theta_kZ_k$).

As $Z_k$ increases, the maps become more and more like mirror images
of one another, which reflects the fact that the offset, $C_1$,
becomes negligible at sufficiently short spacing. The duplication and
reversal maps then become:
$$
Z_{k+1} \simeq \pm f(Z_k)
\quad,
\eqno(6.6)
$$
which is an equality at homoclinicity.

The reduction of the polarity-reversing
case is further enriched when the
symmetry of the system is broken, and the pulses and antipulses are
not simply mirror images of each other. This occurs, for example, when
the nonlinear term in the differential equation is a combination of
$x^2\ {\rm and} \ x^3$.  We shall report on this problem elsewhere.
In its blend of different kinds of individual orbits it is reminiscent
of Kopell and Howard's [23] work on chemically generated patterns.

\bigskip
\goodbreak
{\bf 7. Second-order perturbation theory.}

{\bf 7.1 Preliminaries.}
We now proceed to the derivation of second-order corrections to the
leading-order theory.  For the sake of brevity we fix $\mu=\mu_0$, and
ignore the modifications necessary to treat trains of both pulses and
antipulses.  We also expand $c$ as $c=c_0+\ve c_1 + \ve^2 c_2$, which
makes no modification in the problem at $O(\ve)$. However, before
going on to the next order, we need to solve equation (5.4).

In deriving the timing map (5.13), we made use of the fact that the
operator, $\cl$, contained terms that were localized to the positions
of the pulses. This enabled us to consider the simpler, local
operator, $\cl_k$, and its adjoint's null vector, $N_k$, in order to
construct the solvability condition. Such a procedure resembles
the tight-binding approximation of solid-state physics [10].

In computing the residual $\clr$, we must distinguish between the
two cases, $n=2$ and $n=3$, for the following reason. The right-hand
side of equation (5.4) contains the terms, $nH_k^{n-1}H_{k\pm1}$.
Provided $n>2$, these terms are again localized to the position of the
$k^{th}$ pulse. The inhomogeneous term introduces a localized
particular integral to complement the localized homogeneous solution.
The tight-binding approximation therefore still applies and we can
continue on to order $\ve^2$ accordingly.

For the quadratic case, the inhomogeneous forcing term is no longer
concentrated to the pulse position; the decay in the amplitude of the
$k^{th}$ pulse is equally balanced by the rise of its neighbors in the
product $H_kH_{k\pm1}$.  The forcing term is therefore spread over the
domain covered by the $k^{th}$ pulse, as well as its predecessor and
its successor.  As a result, the calculations of higher-order terms is
made more complicated since the function $\clr$ is not tightly bound
(see  Fig. 20b of Appendix A).  We avoid these complications and
carry out the calculations only for the case $n=3$.

For the cubic case, the correction term $\clr$ is a localized function,
and can be conveniently developed into an expansion of the form,
$$
\clr = \sum_{k=1}^K R_k + \ve \cls
\quad.
\eqno(7.1)
$$
where $R_k=R_k(t-\tau_k)$.
If we use the expansion (7.1) in equation (5.4), then we find that the
leading-order terms satisfy the relation,
$$
 \cl_k R_k = c_1 H_k -
{n\over\ve} (H_k)^{n-1} [H_{k+1}+H_{k-1}]
\quad.
\eqno(7.2)
$$
The timing map discussed in Section 5 is the solvability
condition on this equation, and it assures a finite solution,
$$
R_k = P_k + \phi_k \d H_k
\quad,
\eqno(7.3)
$$
where $P_k$ is a particular solution (and so depends on the
spacings $\Delta_k$ and $\Delta_{k+1}$). The amplitudes,
$\phi_k$, of the homogeneous solutions are yet to be determined,
but this set of arbitrary constants can be interpreted slightly
differently. The ansatz for the pulse train is
$$
x(t) \sim \sum_{k=1}^K [H(t-\tau_k)+\ve \phi_k \dot H (t-\tau_k)
+\ve P_k(t-\tau_k)] + O(\ve^2)
\quad,
\eqno(7.4)
$$
which can be written as
$$
x(t) \sim \sum_{k=1}^K [H(t-\tau_k+\ve \phi_k)
+\ve P_k(t-\tau_k)] + O(\ve^2)
\quad.
\eqno(7.5)
$$
Consequently, the arbitrary constants can be interpreted as
corrections to the positions of the pulses, or {\sl phase
corrections}, and the second-order theory is really about determining
them. We could have achieved this result by letting the $\tau_k$
depend on $\ve$ in the first place.  Then the
$\phi_k$ would have been arbitrary and we would have had to invoke
some other principle to choose them, as is done in some versions of
singular perturbation theory.

\goodbreak
\b
{\bf 7.2 Second-order solvability.}
At $O(\ve^2)$, we find an inhomogeneous ODE for the function
$\cls$. The terms on the right side of this equation are
again tightly bound to the pulse positions, and so we expect (and have
numerically verified -- see Fig. 20a of
Appendix A) that $\cls$ has the same structure as $\clr$.
Hence we seek an approximation of the form
$$
{\cal S} = \sum_{k=1}^K S_k(t-\tau_k) + O(\ve)
\quad .
\eqno(7.6)
$$

The equation for $S_k=S(t-\tau_k)$ is
$$
 \cl_kS_k = \cln_k
\quad,
\eqno(7.7)
$$
where the inhomogeneous term,
$$
 \eqalign{
 \cln_k &= c_2 H_k  + c_1 R_k -\cr
{3\over\ve} H_k [ H_k (R_{k+1}+R_{k-1} + H_{k+2}+H_{k-2})
 &+ R_k (R_k + 2H_{k+1}+2H_{k-1}) + 2H_{k+1} H_{k-1}] \quad.\cr} \
\eqno{(7.8)}
$$
The solvability condition is
$$
\int_{-\infty}^\infty N_k \cln_k dt = 0
\quad.
\eqno(7.9)
$$
After a series of elementary reductions, this condition can be
reduced to the relation,
$$
c_2 =  Q_k - \Phi_{k+1} F'(-\Delta_{k+1}) + \Phi_k F'(\Delta_k) +
{1\over \ve} [F(-\Delta_{k+2}-\Delta_{k+1}) + F(\Delta_k+\Delta_{k-1})]
\quad,
\eqno(7.10)
$$
where
$$
\Phi_k = \phi_k - \phi_{k-1}
\quad,
\eqno(7.11)
$$
$F'(\Delta)=dF/d\Delta$
and $Q_k$ is a rather complicated integral functional which is written
out in Appendix B.

\goodbreak
\b
{\bf 7.3 Phase-correction maps.}
Equation (7.10) defines a map between the phase corrections, $\Phi_k$
and $\Phi_{k+1}$. It formally depends on the values of the four
spacings, $\Delta_{k+2}$, $\Delta_{k+1}$, $\Delta_k$ and
$\Delta_{k-1}$.  The particular spacings, $\Delta_{k+2}$ and
$\Delta_{k+1}$, however, can be uniquely prescribed as functions of
$\Delta_k$ using (5.13).  Therefore, the map (7.10) is of the form,
$$
\Phi_{k+1} = g(\Delta_k) \Phi_k + G(\Delta_{k-1};\Delta_k)
\quad,
\eqno(7.12)
$$
for some functions $g$ and $G$.

In general, the spacing $\Delta_{k-1}$ cannot be uniquely expressed as
a function of $\Delta_k$ because $F(\Delta)$ cannot be inverted uniquely
for positive argument, or, equivalently, because one cannot unambiguously
iterate the spacing map (5.13) backwards. For each value of
$\Delta_k$, there are several values for $\Delta_{k-1}$ that could
precede it. The multiplicity, $M_k$, of such values depends on the
value of $\Delta_k$. But formally, at least, we can write,
$$
\Delta_{k-1} = \Delta_{k-1}^{(i)} (\Delta_k) \quad,
\qquad i=1,\dots,M_k
\quad,
\eqno(7.13)
$$
and also,
$$
G(\Delta_{k-1};\Delta_k) = G^{(i)} (\Delta_k) \quad,
\qquad i=1,\dots,M_k
\quad.
\eqno(7.14)
$$
We can therefore write the map in the form,
$$
\Phi_{k+1} = g(\Delta_k) \Phi_k + G^{(i)}(\Delta_k)
\quad,
\qquad i=1,\dots,M_k
\quad,
\eqno(7.15)
$$
which illustrates how, at second-order, the map becomes $M_k$-valued.

The multiplicity of the map at second order reveals the signature of
higher dimension. In fact, as one goes to yet higher order, the
multiplicity increases in factors of $M_k$, indicating a developing
Cantorial structure.  This reflects how the true timing map is a map
of the plane, but in the limit $\ve\rightarrow 0$, it becomes
contracted to one dimension. The higher-order corrections unfold the
richer nature of the map from the leading-order, degenerate
approximation.  We omit the details of these maps, but they do accord
with the results of numerical experiments, as briefly mentioned in
Appendix A.

Although we have explicitly derived the second-order corrections for
the cubic, and have uncovered asymptotically
the hidden fractal structure for this
case alone, the timing maps for the quadratic case also have this form
(see \S 9.2 and Fig. 17).

\bigskip
\goodbreak
{\bf 8. Comparison with numerical experiment.}

{\bf 8.1 Sample pulse trains.}
The pulse trains shown in Fig. 1 originate from a point on the
unstable manifold at $\bfx=\bfz$, and terminate after a finite number
of pulses.  Since the trains begin from the trajectory that asymptotes
to the origin as $t\rightarrow-\infty$, the asymptotic theory predicts
that the first spacing satisfies,
$$
f(-\Delta_1) = Z_1 = C_1
\quad,
\eqno(8.1)
$$
which is indeed close to the value measured from the numerical
solution of the ODE.  From then on the spacings evolve in more
complicated manner.  Figs. 10 and 11 display the maps corresponding
to the trains of Fig. 1. Illustrated is the behavior of $\Delta_k$
and $Z_k$ as they iterate through successive values.  Points denote
coordinates derived from numerically computed pulse trains, whereas
the curves outline the asymptotic maps (5.13) and (5.15), or
(6.2) and (6.3).  In each case, the two coincide to better than the
resolution of the figure.

For the pulse train in the quadratic case, the trajectory eventually
finds its way around the stable manifold and diverges.  After the
final pulse of this train, the iteration also terminates because it
leads to a negative branch of the map, implying that the following
pulse spacing does not exist (if $Z_{k+1}<0$, then $\Delta_{k+1}$
cannot be real).  Thus the divergence of the ODE is reproduced by the
asymptotic timing map.

Similarly, the duplication-reversal map combination captures the
polarity-reversing property of the pulse train generated by the cubic
equation (Fig. 11). At the pulse preceding the first flip in
polarity, the map iteration reaches a negative
branch of the duplication map and follows into the negative
$\Theta_kZ_k-$plane onto the reversal map. (The catastrophe-inducing
property of the map's convexity is also apparent in this figure;
spacings gradually climb down the spacing map, or spiral outwards in
the $Z_k$-map, as iteration proceeds.)

\medskip
{\bf 8.2 Empirical spacing maps.}
More extensive comparisons of the ODE and the timing map are provided
in Figs. 12 and 13, which show plots of $\Delta_{k+1}$ against
$\Delta_{k}$ over a wide range of spacings.  Again, the points represent
numerically calculated spacings, and the curves show the values from the
maps (5.13) or (6.2).  Because pulse trains do not sample evenly the range
of spacings and the solutions frequently diverge, the empirical points
in Fig. 11 were accumulated from many different trains constructed
using initial conditions,
$$
\bfx = \alpha\; \bfxi_1
\quad,
\eqno(8.2)
$$
with a set of values for the amplitude, $\alpha$, ranging
from $10^{-4}$ to 0.5.

In each figure, four panels are shown, each computed with different
parameter values. The spacings are plotted for the parameter values
listed in Table 1. These choices select locations on the principal
homoclinic loci that straddle the Shil'nikov line, $\delta=1$, and so
the figures compare theory and numerical experiment over an important
range of parameter values.

Because certain spacings immediately cause the ODE to diverge, or
equivalently, always imply imaginary values for $\Delta_{k+1}$, the
quadratic maps shown in Fig. 12 contain extensive gaps. Such
lacunae are not present in the cubic maps since adjacent spacings
can be defined over the entire interval in $\Delta_k$. In both cases,
the empirical and theoretical maps agree for spacings exceeding the
intrinsic pulse width (about 10 for these parameter values).

At shorter spacings, as is particularly apparent for the cubic case,
the theory begins to break down. The empirical maps begin to differ
from the theory in shape and by becoming multivalued. Examination of
the pulses with these spacings reveals why this is the case.  The
trajectory, rather than locking into a path near the principal
homoclinic orbit, tracks a different basic homoclinic orbit as it
emerges from the region around the origin.  In Fig. 13, the
distinct, second curve that appears in panels (a) and (b) is related
to a doubly-peaked homoclinic pulse with polarity $(\up \do)$.
(This pulse is also visible in Fig. 1(c) just beyond $t=250$.)

There are several more curves evident in panel (c), and agreement here
looks somewhat worse.  In actual fact, the multivaluedness is not quite as
bad as it appears graphically, because when multiply peaked pulses
contaminate the train, the numerical algorithm used to locate the ODEs
pulses misidentifies spacings. This breaks up the true curves and
places the pieces in misleading parts of the picture. However,
the breakdown of the asymptotic theory shown in Fig. 13 accords with
our suggestion that the theory is limited to spacings larger than the
intrinsic pulse width.

\goodbreak
\bigskip
{\bf 9. Invariant sets.}

{\bf 9.1 Construction.}
To the left of the homoclinic locus in the $\mu-c$ plane, pulse trains
diverge quickly, and to the right, trains with constant spacing
prevail.  Close to the Shil'nikov curve, $\delta=1$, trains form with
chaotically spaced pulses, and the power-law envelope of the map is
sufficiently concave to prevent the separations from collapsing
immediately. In this parameter regime, one can find pulse trains of
arbitrary length.  There is no evident way to anticipate where such
solutions would lie {\it a priori} just from solving the ODE numerically.
However, with the information contained in the map, we can identify
certain parameter ranges for which trains continue indefinitely. This
amounts to finding sets of points that are invariant under the action of
the map. These {\sl invariant sets} correspond to (occasionally strange)
attractors in the dynamical flow.

An example of an invariant set for $n=2$ is shown in Fig. 14.  The
location of the set and how it can be constructed by perturbing
parameters away from a known map are portrayed in the figure.
Also shown is a typical evolution of the pulse spacings
in the set. Evidently, the
invariant set is roughly characterized by a periodic orbit with
spacing, $\Delta_0$, satisfying,
$$
C_1 = F(\Delta_0) + F(-\Delta_0) =  f(Z_0) - Z_0
\quad,
\eqno(9.1)
$$
and
$$
F'(\Delta_0) = f'(Z_0) = 0
\quad.
\eqno(9.2)
$$
We can perturb the actual solution away from this periodic orbit,
$$
Z_k = Z_0 + \zeta_k
\quad,
\eqno(9.3)
$$
where $|\zeta_k| \ll Z_0$, and introduce this expression into
the map (5.13). On expanding the function $f$, and using the relations
(9.1) and (9.2), we find
$$
\zeta_{k+1} = {1\over2} \zeta_k^2 f''(Z_0) + O(\zeta_k^3)
\quad.
\eqno(9.4)
$$

This short calculation indicates why the invariant part of the map
resembles a slightly skewed parabolic map. Its convexity ensures
chaotic iteration and this is verified by the positivity of the
Lyapunov exponent which is shown in the fourth panel.  But such
iteration is not the only possibility for evolution within an
invariant set. By analogy with standard quadratic maps, we expect
stable fixed points and limit cycles, period doubling cascades and
intermittency, and we have verified the existence of these phenomena
in the ODE.

This kind of invariant set is not the only possible attractor existing
in near-homoclinic conditions. Another example for the cubic equation
is shown in Fig. 15. This set has slightly more structure but
contains only pulses. A sample evolution sequence is shown in panel
(b), and this leads to the Lyapunov exponents shown in panel (c).
Fig. 16 shows an invariant set in which polarity reversals also
occur.  (The pictures of the invariant sets all show spacing maps,
which are double-valued for the cubic sets, but these are visually more
compact than the $Z_k$-sets.)  The second panel of Fig. 16 shows how
the spacings evolve once more, but panel (c) shows a
Poincar\'e section taken at the peaks of the pulses.

It is evident from these three examples that invariant sets are
typically of two kinds. The first case encompasses an extremum of the
spacing map, and it then resembles a quadratic map. The second case
has a richer structure and incorporates part, if not all, of the
$\Delta\rightarrow\infty$ asymptote of the map.

The sets computed for the cubic equation are both chaotic (as shown by
the Lyapunov exponents in Fig. 15), but both are computed for
$\delta>1$, illustrating how the Shil'nikov condition is not always
useful in predicting chaotic dynamics.  A striking feature of the
Lyapunov calculation is the agreement between the leading exponent
computed from the ODE and that derived on iterating the map. In
retrospect, since the map successfully describes the strange attractor
on which the trajectory wanders, the agreement is not surprising,
although there are no theoretical grounds on which this is expected. A
virtue of this result is that the evaluation of the leading Lyapunov
exponent can be achieved with the map, and this is substantially
quicker than computing it from the ODE. For related reasons, the
Floquet exponents of the periodic orbits comprising the strange
attracting set are also well approximated by the asymptotic timing
map. This is a potentially more powerful result because the Lyapunov
exponents are really complicated global averages of various Floquet
exponents over the invariant set.

\medskip
\goodbreak
{\bf 9.2 Fractal structure.}
Localization to an invariant set substantially simplifies the dynamics
of the pulse train.  It limits the portion of the map visited by the
ODE and drastically reduces the number of periodic orbits and fixed
points.  This permits us to compute Lyapunov exponents (see Figs. 14
and 15), examine Poincar\'e sections in detail (see Fig. 16 and Fig. 21
of Appendix A). We can also carry out periodic-orbit expansions
to study the statistical mechanics of the chaotic flow as described in [4].
That application of the method will be taken up elsewhere.

A specific example of the utility of the invariant set is given in
Fig. 17. For sufficiently long pulse trains the invariant interval
is sampled densely enough to build up a picture of the actual error in
the asymptotic maps (5.13) and (6.2). In the three panels of Fig. 17 this is
plotted as a function of spacing for the invariant sets shown in
Figs. 14-16.  Inset panels show greater details, specifically
multiple branches.

The multivaluedness of the error reflects the fractal structure of the
true timing map. In both the quadratic and cubic cases, the error
qualitatively follows the predictions of \S 5.6, in spite of the fact
that the second-order theory applies strictly to $n=3$
(further numerical calculations for the cubic
also show quantitative agreement).
This is
particularly apparent in the first panel for the quadratic set, where
the two branches of the error (extracted by additionally subtracting
off a smooth curve) are compared with the function
$\Delta_{k-1}(\Delta_k)$. In the second of the cubic sets shown in
panel (c), the portion of the error which is magnified further clearly
reveals fractal structure.

\goodbreak
\bigskip
{\bf 10. The incompressible limit.}

{\bf 10.1 Asymptotic re-ordering and reconstitution.}
The asymptotic theory developed in section 5 is tailored for
conditions close to the Shil'nikov condition, $\delta=1$. As we move
away from this vicinity of parameter space, in particular as
$\mu\rightarrow0$, the approximations we have used are no longer
appropriate. For the principal homoclinic loci illustrated in Figs.
2 and 4, at the point where the loci cross the $\mu-$axis, we have
$\delta=1/2$.  In this incompressible limit, pulses therefore decay
asymmetrically to the left and right. Consequently, we have
$$
H(-\bar{\Delta }) \sim [H(\bar{\Delta})]^{1/2} \sim \ve
\quad,
\eqno(10.1)
$$
which violates the ordering of section 3.
This indicates that, at the position of the
$k^{th}$ pulse, the contribution of the
preceding pulse, $H_{k-1}$, is order
$\ve^{-1/2}$ larger than that of the successor.
If we try to account for the asymmetry by expanding all quantities
in series of $\ve^{1/2}$, solvability requires,
$$
C_1 = F(\Delta_k)
\quad,
\eqno(10.2)
$$
implying pulse trains with certain, constant spacings. These trains
compose highly unstable trajectories in phase space, and spacings
typically migrate rapidly away to a spacing catastrophe.  To fix this
problem, more than a small correction is needed.

A preliminary measure for avoiding the breakdown of the asymptotic method
is to reconstitute the two leading-order solvability conditions into a
single map. That reconstituted map is
$$
\tilde C = F(\tilde \Delta_k) +
F(-\tilde \Delta_{k+1}) + F(\Delta+\Delta_{k-1}) - \ve q_k
\quad,
\eqno(10.3)
$$
where
$$
\tilde C = C_1 + \ve c_2
\quad,
\eqno(10.4)
$$
$$
\tilde \Delta_k = \Delta_k + \ve \Phi_k
\quad,
\eqno(10.5)
$$
$$
 q_k = \left[ Q_k \right]_{\Delta_{k+1}=\infty,P_{k+1}=0}
\eqno(10.6)
$$
and by $P_k$ we now mean
$$
P_k = \cl_k^{-1} (C_1 H_k - n H_k^{n-1} H_{k-1})
\quad.
\eqno(10.7)
$$

This three-term recurrence relation is accurate to order $\ve^3$, and
spacings need no longer satisfy equation (10.2). In terms of the
iterative variable, $Z_k=-F^{-1}(-\tilde\Delta_k)$, the reconstituted
map is of the form,
$$
Z_{k+1} =  f(Z_k) - \tilde C + \ve w(Z_k,Z_{k-1})
\quad,
\eqno(10.8)
$$
where
$$
w(Z_k,Z_{k-1}) \sim
{1\over\ve} f(Z_kZ_{k-1}) - q_k
\quad.
\eqno(10.9)
$$

The need to reconstitute the leading asymptotic orders into a map of
the plane, reflects the fact that the dynamical flow in phase space
can no longer be adequately represented by a single coordinate when
$\mu$ becomes small. This is because that flow is not especially
volume-contracting for $\delta< 1$ and, in the incompressible limit,
it becomes volume-preserving (when $\mu<0$ it even expands volumes).
Thus sections through the flow are inevitably two-dimensional.

In spite of the improvements afforded by the reconstitution, we have
not used it in any quantitative fashion, since it cannot capture the
area-preserving property of the spacing map appropriate to $\mu=0$.
That may require the uncovering of an integral of the motion that we
do not yet know analytically. Moreover, in this limit, certain of the
inhomogeneous terms appearing at $O(\ve)$ lose localization to the
pulse positions even for the cubic equation. The second-order theory
is then no longer strictly valid and reconstitution cannot work. The
reconstituted map does, however, qualitatively explain certain features of
the invariant sets for $\mu\rightarrow 0$, as we see next.

\goodbreak
\medskip
{\bf 10.2 Invariant sets and the H\'enon map.}
The breakdown of the one-dimensional asymptotic map and the unfolding
of higher dimension is seen most clearly in sets of spacings that are
invariant under the map.  Consider a periodic orbit with period $\Pi$
close to the homoclinic locus. When defined in terms of period,
this orbit possesses the value of the
parameter $c$ as an eigenvalue. We can therefore
define a function, $c(\Pi)$,
with $c(\infty)=c_0$ (see Fig. 8).  Close to the extrema of this function,
invariant sets always exist and one need not then rely on
the construction shown in Fig. 14 to find them.

Spacing sets for $\mu=0.1$ and $\mu=0$ are shown in the two panels of
Fig. 18 and in Fig. 19.  The weakly dissipative case shown in
Fig. 18a, with $\mu=0.1$, has a small number of distinct branches.
These develop into parabolic-shaped arcs, and other branches emerge as
$\mu$ is decreased still further. Eventually, the set breaks up into
stable spiral foci, periodic arrays of points and limit cycles. As
$\mu$ is then decreased to zero, one finds incompressible sets of the
types shown in Figs. 18b and 19.

Although the iteration has remained in the vicinity of spacing $\approx
19.3$, the spacing set of Fig. 18b is not truly invariant; the
iteration eventually wanders away, revealing this set
to be a weak repeller.  In the second example, several invariant sets are
shown, iteration sequences being marked by the stars.
These sets are periodic and are generated with identical parameter
values, but different initial conditions. Surrounding them are points
that display iterations which eventually wander away
and diverge.  These incompressible sets look like the iterates of a
two-dimensional Hamiltonian map; there are regions of closed orbits
surrounding fixed points, and separatrices adjoining them to chaotic
regions.

As illustrated by the inset panels in Fig. 18, the structures of the
sets, and their intrinsic dependence on the dissipation parameter,
$\mu$, are very reminiscent of the H\'enon map [19].  Why this is the
case can be seen from the expansion (9.3): $Z_k=Z_0+\zeta_k$. If we
fix $\zeta_k$ to be order $\ve$, then the reconstituted map reduces to
$$
 \eqalign{
 \zeta_{k+1} = \ve w(Z_0,Z_0) &+ {1\over 2} \zeta_k^2 f''(Z_0)
 + \ve \zeta_{k-1} {\partial \over\partial Z_{k-1}} w(Z_0,Z_0)\cr
 &\equiv w_0 + {1\over 2} \zeta_k^2 f_0'' + w_k\zeta_k+w_{k-1} \zeta_{k-1}
 \,\cr }
\eqno{(10.10)}
$$
to order $\ve^3$. Equation (10.10) can be rewritten as
the H\'enon map,
$$
y_{k+1} = 1-ay_k^2 - b y_{k-1}
\quad,
\eqno(10.11)
$$
with
$$
y_k = {1\over D}(\zeta_k + w_k/f_0'')
\quad,
\eqno(10.12)
$$
$$
D = {1\over f_0''} (w_0 f_0'' + w_k - w_kw_{k-1} - w_k^2/2)
\quad,
\eqno(10.13)
$$
$$
a = {1\over 2} Df_0''
\eqno(10.14)
$$
and
$$
b = -w_{k-1}
\quad.
\eqno(10.15)
$$
Although $b= 1$ formally gives the area-preserving H\'enon map, this
does not in general correspond to the case $\mu=0$, in line with our
earlier remark that the reconstituted map has only some qualitative
value. In fact, equation (10.12) is derived under the explicit
assumption that $b\sim \ve$.  Despite the uncertainties, we regard it
as worth pursuing these issues because of the interest attaching to
incompressible flows and because the limit $\mu=0$ corresponds to the
much-studied Kuramoto-Sivashinsky equation.

\goodbreak
\bigskip
{\bf 11. Conclusions.}
In this paper we have given a detailed study of the phenomenology of
pulse trains in a dynamical system corresponding to the most
degenerate of triple bifurcations.  These pulse trains occur in
near-homoclinic conditions, which connects our study with
Shil'nikov's theory.  As we have
discussed, there is a great variety of possible homoclinic connections
in these systems.  Each of the different kinds of homoclinic orbit can
be used as building blocks for a pulse train, although we chose only
the principal solutions to work with.

We have sketched mainly circumstances in which the asymptotic
description of pulse trains is successful.  However, we also described
the difficulties still remaining in the incompressible limit in
section 10.  We may mention another circumstance in which
complications pose an interesting unsolved problem. This is suggested
by the homoclinic loci pictured in Figs. 2 and 3. There it is evident
that the various homoclinic branches are comparatively well separated,
except near certain heteroclinic points. Near these points, one cannot
construct pulse trains of one kind of homoclinic orbit without severe
contamination (in the words of section 7) from the neighboring orbits.
This indicates that the phase space is rather more complicated near
the heteroclinic points.

Where the pulse theory works well (close to a single homoclinic locus
and near $\delta=1$) asymptotics gives a description of homoclinic
dynamics. Surprisingly, except for certain invariant spacing sets, the
generic behavior of pulse trains is either to diverge or to terminate
in a spacing catastrophe for $\delta<1$. This indicates that,
at least in the current system, the strange set that temporarily
captures trajectories near the homoclinic orbit is rarely
asymptotically stable.  Moreover, the invariant sets are fragile
because modest changes in parameters destroy them and their basins
of attraction are small.

%To maintain the kind of solution we describe for long times, we need
%a reinjection mechanism that returns the system near to the origin,
%though such a mechanism will bring its own features with it.

The tendency of the systems of type (1.1) to drift away from the
nearly homoclinic orbits explains why numerically one does not find
robust chaos in the conditions where Shil'nikov's theorem implies that
infinitely many unstable periodic orbits densely populate the phase
space surrounding a homoclinic orbit. One cannot therefore always
anticipate the existence of homoclinic chaos under conditions
satisfying Shil'nikov's criterion, $\delta<1$, as was previously
remarked in the heteroclinic context by Howard and Krishnamurti [20].

For a similar reason, the conditions necessary for the formation of
pulse-train solutions to equation (1.1) are relatively restrictive.
The parameters $c$ and $\mu$ must equal their homoclinic values,
$(c_0,\mu_0),$ to within a percent or less. Practically this seems to
suggest that dynamical systems like (1.1) should rarely produce
pulses. Yet we frequently see pulses in experimental situations.  We
suggest that this is so because there is some self-corrective
mechanism forcing those situations into near-homoclinic conditions.
For travelling waves of a partial differential equation
like (1.2), the wave speed $c$ is really
a nonlinear eigenvalue,
%({\it c.f.} Elphick {\it et al.}, 1991)
and therefore the coherent structures may indeed have this
self-correcting property. In systems governed purely by an ODE, the
mechanism creating homoclinic conditions is much less clear.  In
systems that have evolved to some optimal state in time, there may be
good reasons for this.

Whatever the reason for the prevalence of pulse trains, when
conditions are favorable, the method we have described here works
well, as illustrated by the various examples we have presented.  The
lesson of these examples is that, for quantitative study of the
flow generated by such ODEs, the map is both accurate and extremely
useful.  Together with the homoclinic solution, it gives a convenient
description of that flow.  In future work we hope to exploit this
property of the theory in other situations.

\medskip
{\bf Acknowledgments.} We thank Charles Tresser for many
informative discussions,
and R. Worthing and C. Elphick for carefully reading the
manuscript.
Part of this work was carried out at the
1992 session of the GFD summer school, Woods Hole Oceanographic Institute.
Alex Nepomnyashchy offered some references.

\goodbreak
\bigskip
{\bf Appendix A.  Numerical methods and accuracy.}
Equation (1.1) can be solved as either an initial-value problem or a
boundary-value problem.  In the first case, we employ an adaptive
step-size, finite-difference integration algorithm of the
Bulirsch-Stoer (Richardson extrapolation) type [30].  Local error
control on the order of machine precision ($10^{-13}$) is easily
achieved.

For the boundary-value problem, we use a Chebyshev polynomial
expansion for the dependent variable, $x(t)$; specifically
$$
x(t) = \sum_{j=0}^J x_j T_j(t)
\quad.
\eqno(A.1)
$$
Truncation error is kept at a negligible level by choosing a suitable
number of polynomials, $J$; typically 64 for solutions on a small domain,
up to 128 for pulses approaching homoclinicity. For periodic
solutions we require
$$
\eqalign{
x(-L) &= x(L) \cr
\dot x(-L) &= \dot x(L) = 0 \cr
\ddot x(-L) &= \ddot x(L). \cr
}\eqno(A.2)
$$
(The second condition fixes the phase.)  We use the Lanczos
tau method to enforce (A.2).  By aligning the central peak with the
ends of the interval $[-L,L]$, as the domain size is progressively
increased in the direction of a homoclinic solution we profit from the
high boundary layer resolution characteristic of a Chebyshev
expansion.  Solutions for the $\{ x_j \}$ are found by Newton's
method with the proviso that the simultaneous determination of the
eigenvalue, $c$, requires an auxiliary equation.  There is no unique
prescription for this. We choose to form an integral constraint
obtained by multiplying (1.1) by $x$ and integrating over one period.

The departure of $c$ from the homoclinic value scales approximately as
$\exp(-\gamma L)$ times a characteristic value of $F$ (see Fig. 8).
We use a domain of size $L=40\gamma$, which can be regarded as
effectively infinite. Then (A.1) gives an expansion for $H(t)$ in
the vicinity of $t=0$.  This can be extended to arbitrarily large $t$
by patching an asymptotic tail onto the solution, and it is then
straightforward to compute the integral function, $F(\Delta)$.

The Chebyshev expansion (A.1) also permits us easily to construct the
higher-order corrections $R_k$ and $S_k$.  The particular
solution is generated by inverting the operator $\cl_k$ using a
singular-value decomposition, discarding the contribution from the
nearly zero eigenvalue (typically its value is of order $10^{-10}$).

Examples of the functions $H_k$, $R_k$ and $S_k$
for a periodic solution are shown in
Fig. 20(a). These were computed for $\mu = 1/\sqrt 3$
and $L=\Delta=19$. The solutions
are all strongly concentrated at $t=0$, verifying the accuracy of the
tight-binding approximation. This is to be contrasted with Fig.
20(b) which shows $H_{k-1}+H_k+H_{k+1}$ and $R_k$ for a periodic orbit
of the quadratic with $\Delta=L=18$. The
residual here is clearly spread over the interval between
$\tau_{k-1}$ and $\tau_{k+1}$ as described in section 7.

Another test of the ansatz (5.2) is had by comparing the prediction of
$\epsilon c_1 = \epsilon [F(L) + F(-L)]$ with the computed difference
$c-c_0$. Some results for the quadratic equation are shown in Fig. 8.
For the cubic equation, with
$\mu=1/\sqrt{3}$, we find the eigenvalue corrections listed
in Table 2, for three different periodic orbits.  The first entries
of the table are $c-c_0$ and $\ve c_1$ which agree to the third
significant figure. The difference is then listed as the second entry
and compared to $\ve^2 c_2$, calculated according to the method
outlined in the main body of the paper (section 7). For periods greater
than those shown in Table 2,
the correction, $\ve^2 c_2$, while an
increasingly accurate prediction of the asymptotic theory, becomes
inaccessible owing to the finiteness of numerical precision arising
both from machine round-off error and from the use of a patched
asymptotic tail for $H_0$ and $N$.

A possibly more useful test of the asymptotics is a comparison of
predicted and computed Poincar\'e sections for one of the invariant
spacing sets discussed in section 8. To do this, we iterate the map
with a specified choice of $\epsilon c_1$ and generate a table of peak
separations.  For each successive pair of nearest-neighbor spacings,
we compute $R_k$ and the phase corrections, $\Phi_k$. From this we
reconstruct the solution to order $\ve^2$ and tabulate the values of
$x$ and $\ddot x$ at the
peak of the pulse. The result is shown in Fig. 21 for the cubic case
with $\mu = 1/\sqrt{2}$ (the invariant set of Fig. 15). Plotted are
the variations of $x$ and $\ddot x$. The points show the values
extracted form the numerical solution of the ODE, and the curve shows
the reconstructed section.

It should be noted that reconstructing the solution in the fashion we
have outlined is notably more efficient than solving the differential
equation itself; in solving for $R_k$ the operator ${\cal L}_k$ is
fixed and only the inhomogeneous right-hand side of the equation (7.2)
changes. In addition the method is sufficiently accurate that it may,
for many purposes, be considered exact.  This recipe of orbital
reconstruction can then provide explicit solutions much more rapidly
than any explicit numerical integration algorithm.

\goodbreak
\bigskip
{\bf Appendix B.}
In section 7, we detail the derivation of the second-order solvability
condition. That equation contains a rather complicated integral
functional, $Q_k$, which we write down in this appendix:
$$
Q_k =
{3\over \ve I_0} \int_{-\infty}^\infty
   H^2(t) N(t) [P_{k-1}(t+\Delta_k) + P_{k+1}(t-\Delta_{k+1})] dt
$$
$$
+{6\over I_0}\int_{-\infty}^\infty H(t)N(t)
H(t+\Delta_k)H(t-\Delta_{k+1} )dt
$$
$$
+{6\over I_0}\int_{-\infty}^\infty H(t)N(t)P_k(t)
[H(t+\Delta_k)+H(t-\Delta_{k+1})]dt
$$
$$
+{3\over I_0}\int_{-\infty}^\infty H(t)N(t) P_k^2(t) dt
-{c_1\over I_0}\int_{-\infty}^\infty N(t) P_k(t) dt
\quad.
\eqno(B.1)
$$
$Q_k$ is consequently dependent upon the four spacings
$\Delta_{k+2}$,
$\Delta_{k+1}$,
$\Delta_{k}$ and
$\Delta_{k-1}$.

\vfil
\eject

\c {REFERENCES}

\item {[1]} M.J. Ablowitz and H. Segur, {\sl Solitons and the Inverse
Scattering Transform}, SIAM, Philadelphia, 1981.

\item {[2]} A. Arneodo, P.H. Coullet and E.A. Spiegel,
{\sl Dynamics of triple convection},
Geophys. Astrophys. Fluid Dynamics, {\bf 31} (1985a), pp. 1-48.

\item {[3]} A. Arneodo, P.H. Coullet, E.A. Spiegel and C. Tresser,
{\sl Asymptotic chaos},
Physica D, {\bf 14} (1985b), pp. 327-347.

\item {[4]} R. Artuso, E. Aurell, and P. Cvitanovi\'c,
{\sl Recycling of strange sets: I. Cycle expansions}
Nonlinearity, {\bf 3} (1990), pp. 325-360,
and
{\sl Recycling of strange sets: II. Applications}
Nonlinearity, {\bf 3} (1990), pp. 361-386.

\item {[5]} D.J. Benney,
{\sl Long waves on liquid films},
J. Mathematics and Physics, {\bf 45} (1966), pp. 150-155.

\item {[6]} F.H. Busse and L. Kramer,
{\sl Nonlinear Evolution of Spatio-temporal Structure in
Dissipative Continuous Systems}, Vol. 255 NATO Advanced
Study Institute, Series B: Physics, Plenum, New York, 1990.

\item {[7]} H. Chang, E.A. Demekhim and D.I. Kopelevich,
{\sl Laminarizing effects of dispersion in an active-dissipative
nonlinear medium}, Physica D, {\bf 63} (1993), pp. 299-320.

\item {[8]} P. Coullet and C. Elphick,
{\sl Topological defect dynamics and Melnikov's theory},
Phys. Lett., {\bf 121A} (1987), pp. 233-236.

\item {[9]} A. Einstein, L. Infeld and B. Hoffman,
{\sl The gravitational equations and the problem of motion},
Ann. Math., {\bf 39} (1938), pp. 65-100.

\item {[10]} C. Elphick, E. Meron and E.A. Spiegel,
{\sl Patterns of propagating pulses},
SIAM J. Applied Math., {\bf 50} (1990), pp. 490-503.

\item {[11]} C. Elphick, G.R. Ierley, O. Regev and E.A. Spiegel,
{\sl Interacting localized structures with Galilean invariance},
Phys. Rev. A, {\bf 44} (1991), pp. 1110-1122.

%\item {[11]} A.C. Fowler and C.T. Sparrow,
%{\sl },
%Nonlinearity, {\bf 4} (1991), pp. 1159-.

\item {[12]} A.V. Gaponov-Grekhov, M.I. Rabinovich and J. Engelbrecht,
eds.,  {\sl Nonlinear Waves 1}, Res. Rep. in Phys.
Springer-Verlag, New York, 1989.

\item {[13]} P. Gaspard and G. Nicolis,
{\sl What can we learn from homoclinic orbits in chaotic dynamics?},
J. Stat. Phys., {\bf 31} (1983), pp. 499-518.

\item {[14]} P. Glendinning, {\sl Bifurcations Near Homoclinic
Orbits With Symmetry}, Phys. Lett., {\bf 103A} (1984), pp. 163-166.

\item {[15]} P. Glendinning and C.T. Sparrow,
{\sl Local and global behavior near homoclinic orbits},
J. Stat. Phys., {\bf 35} (1984), pp. 645-696.

\item {[16]} P. Glendinning and C.T. Sparrow,
{\sl T-points: A Co-dimension Two Heteroclinic Bifurcation},
J. Stat. Phys., {\bf 43} (1986), pp. 479-488.

\item {[17]} M.A. Gol'dshtik and V.N. Shtern,
{\sl Theory of structural turbulence},
Sov. Phys. Dokl., {\bf 257} (1981), pp. 1319-1322.

\item {[18]} K.A. Gorshkov and L.A. Ostrovsky,
{\sl Interactions of solitons in non-integrable systems:
direct perturbation method and applications},
Physica D, {\bf 3} (1981), 424-438.

\item {[19]} M. H\'enon,
{\sl A two-dimensional mapping with a strange attractor},
Comm. Math. Phys., {\bf 50} (1976), pp. 69-77.

\item {[20]} L.N. Howard and R. Krishnamurti,
{\sl Large-scale flow in turbulent convection: a mathematical model},
J. Fluid Mech., {\bf 170} (1986), pp. 385-410.

\item {[21]} T. Kawahara and S. Toh,
{\sl Pulse interactions in an unstable dissipative-dispersive nonlinear
system},
Phys. Fluids, {\bf 31} (1987), pp. 2103-2111.

\item {[22]} K. Kawasaki and T. Ohta, {\sl Kink dynamics in
one-dimensional nonlinear systems},
Physica, {\bf 116A}
(1982), pp. 573-593.

\item {[23]} N. Kopell and L.N. Howard,
{\sl Target patterns and horse shoes from a perturbed central-force problem:
some temporally periodic solutions to reaction-diffusion equations},
Studies in Applied Math., {\bf 64} (1981), pp. 1-56.

\item {[24]} Y. Kuramoto,
{\sl Chemical Oscillations, Waves and Turbulence},
Springer, Berlin, 1984.

\item {[25]} R.E. La-Quey, S.M. Mihajan, P.H. Rutherford and W.M. Tang,
{\sl Nonlinear saturation of the trapped-ion mode},
Phys. Rev. Lett., {\bf 34} (1975), pp. 391-396.

%\item {[23]} P. Manneville,
%{\sl Dissipative Structures and Weak Turbulence},
%Academic, New York, 1990.

\item {[26]} D.W. Moore and E.A. Spiegel,
{\sl A thermally excited nonlinear oscillator},
Astrophys. J., {\bf 143} (1966), pp. 871-887.

\item {[27]} L.P. Shil'nikov, {\sl A case of the existence of a
countable number of periodic motions},
Soc. Math. Dokl., {\bf 6} (1965), pp. 163-166.

\item {[28]} L.P. Shil'nikov, {\sl A contribution to the problem of the
structure of an extended neighborhood of a rough equilibrium state of
saddle-focus type},
Math. USSR Sbornik, {\bf 10} (1970), pp. 91-102.

\item {[29]} E.A. Spiegel, {\sl Physics of Convection},
Proceedings of the Summer Study
Program in G.F.D., W. Malkus and F.K.  Mellor, eds., Woods Hole Oceanographic
Inst., WHOI-81-102, 1981, pp. 1-77.

\item {[30]} J. Stoer, and R. Bulirsch, {\sl An Introduction to
Numerical Analysis}, Springer-Verlag, Berlin, 1984.

\item {[31]} C. Tresser, {\sl About some theorems of Shil'nikov},
Ann. Inst. H. Poincar\'e, {\bf 40} (1984), pp. 441-461.

\vfil
\eject
\c {\bf Figure Captions}
\bigskip

\item{F{\ninerm IG.} 1:} Panel (a) shows a pulse train for the
quadratically nonlinear equation. Parameter values are $\mu = 1/\sqrt 2$
and $c=1.92847$. Panel (c) shows a pulse train for the cubic equation,
with $\mu=1/\sqrt 3$ and $c=1.04430$.  In the neighboring panels (b) and
(d), the spacings, $\Delta$,
of the pulses are displayed as functions of the
position of the leading pulse of each pair.

\item{F{\ninerm IG.} 2:} Homoclinic loci in the $\mu-c$ plane for the
quadratically nonlinear equation. The points indicate the locations at
which the phase portraits of Fig. 4 are drawn, and the broken curve
represents the Shil'nikov curve $\delta=1$. The
order, $N$, of homoclinic orbits along the loci is also indicated.

\item{F{\ninerm IG.} 3:}
Homoclinic loci in the $\mu-c$ plane for the cubic equation, plotted in
a similar way to Fig. 2. The order, $N$, and polarity of
homoclinic orbits along the loci is indicated.

\item{F{\ninerm IG.} 4:} A set of example phase portraits computed with
parameter values indicated by points just off the homoclinic loci in
Fig. 2. These portraits are projections onto the plane $(X,Y)$ with
$X=4x-5\dot x$ and $Y=4x+5\dot x$.
The stars indicate the positions of the fixed
points.

\item{F{\ninerm IG.} 5:} A set of example phase portraits taken at the points
just off the homoclinic loci
indicated in Fig. 3. These are projected onto the plane $(X,Y)$ with
$X=4x-11\dot x-29\ddot x/4$ and $Y=4x+3\dot x+13\ddot x/4$. The stars
indicate the positions of the fixed points.

\item{F{\ninerm IG.} 6:} The function, $F(\Delta)$, for the quadratic
equation with $\mu = 1/\sqrt 2$ and $c=1.92847$. The exponentially
decaying behaviour of $F$ to the left, and its envelope to the
right are indicated by the broken and dotted curves.

\item{F{\ninerm IG.} 7:} An example map for the quadratic equation with
$\mu =1/\sqrt 2$ and $c=1.92847$. The first panel shows the map in
spacings, and the second panel (with a magnified image in an inset)
shows the map in $\ve Z_k$. Also shown are the lines with unit slope.
The function $F$ from which this map
is determined is shown in Fig. 6.

\item{F{\ninerm IG.} 8:} The eigenvalue correction, $\ve c_1$, as a
function of $\Delta$ for the quadratic equation with parameter values,
$\mu=1/\sqrt 2$ and $c_0 = 1.928472$. The stars indicate points computed
directly from the ODE and the curve shows $\ve[F(\Delta)+F(-\Delta)]$.
In panel (b), $\log |\ve c_1|$ is plotted against $\Delta$ to show
structure at larger spacings.

\item{F{\ninerm IG.} 9:} The timing map for the cubic equation, computed for
the
parameter values, $\mu=1/\sqrt 3$ and $c_0 = 1.044341$. The ``duplication
map'' is shown by the continuous curve, and the ``reversal map'' by
the broken curve. Panel (a) shows the map in spacing, and panel (b) the
map in $\ve \Theta_kZ_k$.  Also shown are the lines with unit slope.

\item{F{\ninerm IG.} 10:} Map iterations for the example pulse train shown
in Fig. 1(a) and (b). Panel (a) shows iteration in the plane of successive
spacings.  Panel (b) shows the $\ve Z_k-\ve Z_{k+1}$ plane. In both cases, the
curves are drawn using the asymptotic timing map, but the iteration
points are extracted from the numerical solution of the ODE.

\item{F{\ninerm IG.} 11:} Map iterations for the pulse-antipulse train shown
in Fig. 1(c) and (d).  In panel (a), the iteration through successive
spacings is displayed; the continuous curve shows the duplication
map and the broken curve the reversal map. In panels (b) and
(c), the same iteration is shown in the $\ve\Theta_kZ_k$ plane. Here, the
branch to the right is the duplication map, and that to the left is
the reversal map. The curves are drawn using the
asymptotic timing map, but the iteration points are extracted from the
numerical solution of the ODE. Panel (c) is a magnification of the
central parts of the map of panel (b).
 In agreement
 with the divergence found for the ODE, the
 map iteration terminates at the diagonal and cannot be continued to
 produce a successor value of $\Delta_{k+1}$.

\item{F{\ninerm IG.} 12:} Spacing maps for $n=2$. The four panels
correspond to the four sets of parameter values listed in Table 1.
The continuous curves are drawn using the asymptotic timing map, and
the points are ``empirical'' values computed with the ODE.

\item{F{\ninerm IG.} 13:} Spacing maps for $n=3$ and the
parameter values listed in Table 1.
The curves show the asymptotic timing map, with the
duplication map represented by the continuous curve and the reversal
map by the broken curve, and the points are ``empirical'' values
computed with the ODE.

\item{F{\ninerm IG.} 14:} An invariant set for the
quadratically nonlinear equation.  Panel (a) shows the construction of
an invariant piece of the map; the boxed portion of the map is
invariant when vertically displaced as shown, so that
the box's diagonal lies
along the line of unit slope. Panel
(b) shows the resulting spacing set,
with points from the ODE and the curve from the asymptotic timing map.
Panel (c) shows the evolution of spacings (for the ODE)
from an initial point within
the set, and panel (d) shows the calculation of Lyapunov exponents for
both ODE, labelled $\lambda_{ODE}$ (the two largest exponents are displayed,
one of which should asymptote to zero),
and map, labelled $\lambda_{map}$.
The set is calculated with parameter values $\mu =
1/\sqrt 2$ and $\ve c_1 = 1.1\times 10^{-4}$. Shil'nikov's parameter,
$\delta=0.92$.

\item{F{\ninerm IG.} 15:} An invariant set for the cubic equation.
Panel (a) shows the set, with points from the ODE and the curve drawn
from the asymptotic timing map. Panel (b) displays a sample evolution
of spacings (from the ODE)
within the set and panel (c) shows a Lyapunov exponent
calculation (the two ODE exponents are labelled $\lambda_{ODE}$, and that
from the map is $\lambda_{map}$).
Parameter values are $\mu = 1/\sqrt 2$ and $\ve c_1 =
1.3\times 10^{-4}$. The Shil'nikov parameter $\delta = 1.09$.

\item{F{\ninerm IG.} 16:} An invariant set for the cubic equation
containing both pulses and antipulses.  Panel (a) shows the set, with
points from the ODE and the curve drawn from the asymptotic timing
map. Panel (b) displays a sample evolution of spacings
(from the ODE) within the set
and panel (c) shows a Poincar\'e section
(also from the ODE) at the peak of the pulses.
The coordinates of this section are the variations of $|x|$ and $|\ddot
x|$.  Parameter values are $\mu = 0.7$ and $\ve c_1 = 1.1\times
10^{-4}$. The Shil'nikov parameter $\delta = 1.07$.

\item{F{\ninerm IG.} 17:} Error in the map computed for the invariant sets
of Figs. 14 to 16. In panel (a), the inset picture shows the
double-valuedness of the error, revealed on subtracting a smooth
curve, and compared to the function $\Delta_{k-1}(\Delta_k)$, {\it
i.e.} the inverse of the map. In the inset picture of panels (b) and
(c), certain pieces of the multivalued branches of the error are
magnified still further to show the fractal structure.  In panel (c),
finer details are shown rather than the complete error; the asymptotic
map is somewhat less accurate near the divergences of the function
$\Delta_{k+1}(\Delta_k)$.

\item{F{\ninerm IG.} 18:} Spacing sets as $\mu\rightarrow 0$.
Panel (a) shows an invariant set for $\mu=0.1$.  Inset is a picture of
the H\'enon map, (10.11), for $b=-0.2$ and $a=1.85$.  Panel (b) shows
a spacing set for $\mu=0.$ Inset is a picture of the H\'enon map for
$b=-1$ and $a=1.03$.  These latter two examples are not truly invariant
sets since if one were to wait for long enough, the iteration would
eventually wander away and diverge, revealing the object to be a weak
repeller.

\item{F{\ninerm IG.} 19:} Several invariant sets for $\mu=0$. These sets
all correspond to the same parameter values, but are computed from
different initial conditions and are portrayed by the stars.  All
three form periodic curves.  The dense outlying points indicate
iterations that stay near the invariant sets for extended periods of
time whilst wandering through an adjoining chaotic region.

\item{F{\ninerm IG.} 20:} (a) The functions $H_k(t)$, $R_k(t)$ and $S_k(t)$
for a periodic orbit of
the cubic equation with $\mu=1/\sqrt 3$
and period $L = \Delta =19$.  (To facilitate
comparison of structure, the amplitudes of all three have been
normalized to unity.) In panel (b), for a periodic orbit of the
quadratic equation, the three pulses $H_{k-1}$,
$H_{k}$ and $H_{k+1}$, spaced by $\Delta=18$
(and normalized to peak at unity), are compared to the
nonlocal residual $R_k$ (normalised to peak at -1).

\item{F{\ninerm IG.} 21:} Poincar\'e sections for the invariant set shown in
Fig. 15. The coordinates are the variations of $x$ and $\ddot x$.
The section is taken at the
peak of each pulse, and is represented by stars. The curve shows the
section reconstructed from the asymptotic theory.

\vfil
\eject
\noindent{\bf Table 1:}  Parameter values for Figs. 12 and 13.

\vskip .5cm

\hrule

\vskip .25cm

\centerline{$n=2$}

\vskip .5cm

\hrule

\vskip .5cm

\settabs 5\columns
\+ &  $\mu=\mu_0$ & $c_0$ & $\ve c_1$ & $\delta$ \cr
\+ & & & & \cr
\hrule
\+ & & & & \cr
\+ &  $1/\sqrt 2$ & 1.928472 & $1\times 10^{-4}$ & 0.92 \cr
\+ &  $1/\sqrt 2$ & 1.928472 & $2\times 10^{-7}$ & 0.92 \cr
\+ &  1 & 2.250830 & $1\times 10^{-2}$ & 1.08 \cr
\+ &  1 & 2.250830 & $2\times 10^{-4}$ & 1.08 \cr

\vskip .25cm

\hrule

\vskip 1.cm

\centerline{$n=3$}

\vskip .5cm

\hrule

\vskip .5cm

\settabs 5\columns
\+ &  $\mu=\mu_0$ & $c_0$ & $\ve c_1$ & $\delta$ \cr
\+ & & & & \cr
\hrule
\+ & & & & \cr
\+ & $1/\sqrt 3$ & 1.044341 & $5\times 10^{-6}$ & 0.97 \cr
\+ & $1/\sqrt 3$ & 1.044341 & $5\times 10^{-8}$ & 0.97 \cr
\+ & $1/\sqrt 2$ & 1.111617 & $5\times 10^{-5}$ & 1.07 \cr
\+ & $1/\sqrt 2$ & 1.111617 & $8\times 10^{-7}$ & 1.07 \cr

\vskip .25cm

\hrule

\vfil
\eject

\noindent{\bf Table 2:} Eigenvalue corrections for periodic orbits.

\vskip .5cm

\hrule

\vskip .5cm

\settabs 5\columns
\+ & & $\Delta=14$ & $\Delta=17$ & $\Delta=20$ \cr
\+ & & & & \cr
\hrule
\+ & & & & \cr
\+ $c-c_0$           & & $-2.797583\times 10^{-3}$
                       & $ 6.914763\times 10^{-4}$
                       & $-1.094638\times 10^{-5}$  \cr
\+ $\ve c_1$         & & $-2.815155\times 10^{-3}$
                       & $ 6.893459\times 10^{-4}$
                       & $-1.093779\times 10^{-5}$  \cr

\vskip .25cm

\hrule

\vskip .5cm

\+ $c-c_0-\ve c_1$   & & $ 1.757227 \times 10^{-5}$
                       & $ 2.130421 \times 10^{-6}$
                       & $-8.583394 \times 10^{-9}$ \cr
\+ $\ve^2 c_2$       & & $ 1.748241 \times 10^{-5}$
                       & $ 2.033449 \times 10^{-6}$
                       & $-2.486001 \times 10^{-9}$ \cr

\vskip .25cm

\hrule

\end